\documentclass[prb,twocolumn,floatfix,amssymb,showpacs]{revtex4}

\usepackage{graphicx}
\usepackage[intlimits]{amsmath}

\newcommand{\es}{\varepsilon_\mathrm{s}}
\newcommand{\bE }{\boldsymbol{\mathcal{E}}}
\newcommand{\mcE}{\mathcal{E}}
\newcommand{\Hmlt}{\mathcal{H}}

\begin{document}

\title{Stark effect of shallow impurities in Si}

\author{G.~D.~J.~Smit}
\email{g.d.j.smit@tnw.tudelft.nl}
\author{S.~Rogge}
\email{s.rogge@tnw.tudelft.nl}
\author{J.~Caro}
\author{T.~M.~Klapwijk}
\affiliation{Department of NanoScience, Delft University of
Technology, Lorentzweg 1, 2628 CJ Delft, The Netherlands}

\begin{abstract}
We have theoretically studied the effect of an electric field on
the energy levels of shallow donors and acceptors in silicon. An
analysis of the electric field dependence of the lowest energy
states in donors and acceptors is presented, taking the
bandstructure into account. A description as hydrogen-like
impurities was used for accurate computation of energy levels and
lifetimes up to large (several MV/m) electric fields. All results
are discussed in connection with atomic scale electronics and
solid state quantum computation.
\end{abstract}

\date{\today}


\maketitle

\section{Introduction}

The field of atomic scale electronics (ASE) aims at controlling
charge and spin in semiconductors at the level of
\emph{individual} dopant atoms. Such an ability is very
attractive, both for physics and for the development of (quantum)
devices. From a fundamental point of view, dopant atoms are
interesting, because they can be considered as the solid state
analogue of atoms in free space. Several well-known effects from
atomic physics (e.g.\ the Stark effect and Zeeman effect) have
been studied in great detail in large ensembles of dopant atoms
\cite{ramdas81}. The prospect of experimentally realizing atomic
scale electronics has renewed the interest in dopant atoms.
Measurement and control of individual dopant atoms allows for the
observation of quantum coherent time evolution and interactions of
the dopant's wave functions, which is essential for the operation
of a quantum computer.

Manipulation of a single particle's wave functions can be realized
by using a local magnetic or electric field. Such a field can be
used either to perform the desired manipulation itself, or to
provide a local perturbation allowing for addressing a single
impurity by a global radiation field. A local electric field could
be realized by putting a small gate close to a dopant atom, which
is in principle accomplishable with current technology. An
ultimate application of gate-manipulation is found in the solid
state quantum computer as proposed by Kane \cite{kane98,kane00}.

To get more insight in the physics of atomic scale electronic
devices, it is essential to try to predict their potential
behavior. A first step is the description of isolated dopant atoms
in a (homogeneous) electric field. Much more difficult is accurate
modelling of a the time evolution of a dopant atom wave function
in an inhomogeneous field and the description of the interaction
of two or more dopants in a field.

Dopant atoms binding one electron or hole can be described as a
hydrogen atom, where the vacuum values of the dielectric constant
and the electron mass are replaced by the appropriate values for
the semiconductor. This `scaled hydrogen model' (SHM) provides a
reasonable description of the dopant atom's energy levels.
Therefore, it is useful to look at existing studies of the Stark
effect in the hydrogen atom. Calculation of the shift and
splitting of the hydrogen energy levels up to very large electric
field have been carried out by several different methods
\cite{alvarez91,fernandez96,ivanov97}. Within the SHM, these
results can be directly translated to dopant atoms in a uniform
electric field. However, we found that almost no actual results of
such calculations in the (field)range of interest for ASE have
been published.

The SHM also offers a manageable way to describe a dopant atom in
an inhomogeneous electric field. Recently, several calculations
using this framework have been published
\cite{kettle03,smit03,fang02} in the context of quantum computing.
However, the SHM fails in the explanation of effects where it is
essential that the bandstructure of the semiconductor is taken
into account (as an example, see Ref.~\onlinecite{koiller02}).

Many measurements of the energy levels of dopant atoms in
semiconductors (large ensembles) are known, but only a few
concerning the effect of a uniform electric field have been
reported, presumably because such measurements are much more
difficult than e.g.\ measurements in a magnetic field or under
stress. Among them are spectroscopic measurements of the boron
energy levels in silicon subject to electric fields up to
0.15~MV/m \cite{white67}. Electron-spin-resonance experiments
\cite{kopf92} demonstrated that the electric field couples
linearly to the acceptor ground state. The magnitude of the
effective electric dipole moment for linear Stark coupling has
been estimated as 0.26~D for boron acceptors in silicon
($1~\mathrm{D}=3.3 \times 10^{-30}$~Cm). Photo-ionization
measurements have shown a very large electric field effect on the
phosphorus ground state in Si \cite{guichar72}, but this was
measured in highly doped samples where the interaction between
dopants dominates the Stark effect of individual energy levels.
Finally, quadratic level shifts have been observed in deep
selenium double donors in Si, located in the space charge region
of a diode \cite{larsson88}.

In this paper, we will theoretically investigate the effect of a
uniform electric field on isolated shallow impurities in silicon.
Primary interest for ASE will be in the ground state and possibly
the first few excited states. These states are the only ones that
are well separated from neighboring levels and at low temperatures
only the ground state is occupied. Therefore, we focus on the
lowest energy states of impurities in silicon. First, we derive
the shift, splitting and wave functions of the lowest donor levels
in silicon in a small uniform electric field, taking full account
of the multiple valley conduction band structure
(Sec.~\ref{sec:donors}). We briefly outline a similar calculation
for acceptors in silicon (Sec.~\ref{sec:acceptors}). The results
are useful for applications where a local gate is used to bring a
single dopant atom into resonance with a global radiation field
(nuclear magnetic resonance, electron spin resonance). Moreover,
they can be used to outline the limitations of the SHM. Second, in
Sec.~\ref{sec:H} we present accurate numerical calculation of the
Stark effect in silicon within SHM, from zero field up to fields
that are relevant for atomic scale electronics and quantum
computing (several MV/m; see for instance
Ref.~\onlinecite{kane00}). Finally, we conclude by discussing
possible extensions and alternatives of our methods which are
useful to address issues in ASE (Sec.~\ref{sec:dis}).

\section{Donors}

\label{sec:donors}

\subsection{The donor ground state}

Group theory is a powerful tool to derive various properties of
dopant wave functions in a semiconductor. In order to provide the
necessary background and to fix the notation, we will briefly
review some relevant properties of donor levels in silicon (see
e.g.\ Ref.~\onlinecite{kohn55}). Degeneracy due to spin is not
lifted by an electric field in donors. For simplicity, we will
therefore not count those degeneracies in this section.

\begin{figure}
  \centering
  \includegraphics[width=8.6cm]{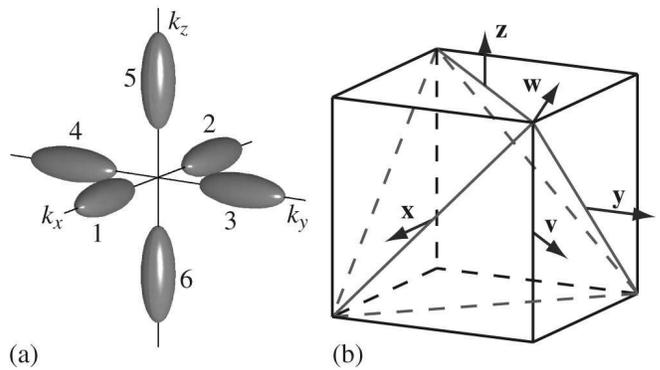}
  \caption{(a) Schematic representation of the conduction band valleys
  of silicon as constant energy surfaces in $k$-space.
  The six valleys are labelled by numbers, e.g.~4
  represents the $[0\bar{1}0]$ valley. (b) Definition of
  the coordinate system with respect to the Si-crystal unit cell. We have
  $\mathbf{x}\parallel [100]$, $\mathbf{y}\parallel [010]$,
  $\mathbf{z}\parallel [001]$, $\mathbf{v}\parallel [110]$, and
  $\mathbf{w}\parallel [111]$. The orientation of the figure in part
  (a) and (b) is the same.}
  \label{fig:cube}
\end{figure}

The conduction band of silicon has six equivalent minima located
on the [100] and equivalent axes. These minima are commonly called
`valleys' and we label them by the numbers 1 to 6 as shown in
Fig.~\ref{fig:cube}(a). The band structure in the vicinity of
valley~1, located in $k$-space at $\mathbf{k}_1=(k_0,0,0)$, can be
approximated as
\begin{equation*}
  E=E_0+\frac{\hbar^2}{2m_\parallel}(k_x-k_0)^2+
    \frac{\hbar^2}{2m_\perp}(k_y^2+k_z^2),
\end{equation*}
where $m_\parallel=0.98m$ and $m_\perp=0.19m$ are the electron
effective masses and $m$ is the free electron mass. Furthermore,
$k_0=0.85\frac{2\pi}{a}$ \cite{feher59}, where $a$ is the size of
the silicon unit cell. Similar expressions hold for the remaining
five valleys.

From effective mass theory (EMT) it follows that the ground state
wave function of the Hamiltonian of an electron bound to a donor
can be written as \cite{luttinger55}
\begin{equation}
  \Psi(\mathbf{r})=
    \sum_{\mu=1}^6 \alpha_\mu F_\mu(\mathbf{r})\varphi_\mu(\mathbf{r}),
  \label{eq:EMTgs}
\end{equation}
where the $\alpha_\mu$ are numerical coefficients and the
$F_\mu(\mathbf{r})$ are envelope wave functions, which are slowly
varying on the length scale of $a$.
$F_1(\mathbf{r})=F_2(\mathbf{r})$ satisfy the hydrogen-like
Schr\"odinger equation
\begin{multline}
  -\Biggl[\frac{\hbar^2}{2m_\parallel}\frac{\partial^2}{\partial x^2}+
  \frac{\hbar^2}{2m_\perp} \left(\frac{\partial^2}{\partial y^2}+
    \frac{\partial^2}{\partial z^2}\right)+ \\
  \frac{e^2}{4\pi\varepsilon r}\Biggr]F(\mathbf{r})= E\,F(\mathbf{r})
  \label{eq:EMTschr}
\end{multline}
and similar equations hold for the remaining $F_\mu$. The
$\varphi_\mu(\mathbf{r})$ are Bloch-wave functions at the minimum
of valley $\mu$ and can be written as $e^{i\mathbf{k}_\mu \cdot
\mathbf{r}}u_\mu(\mathbf{r})$, where $u_\mu(\mathbf{r})$ has the
periodicity of the silicon crystal lattice. Because for all $\mu$
the eigenvalues resulting from Eq.~(\ref{eq:EMTschr}) are the
same, Eq.~(\ref{eq:EMTgs}) shows that the degeneracy of each of
these eigenvalues is multiplied by six for the total wave
functions $\Psi(\mathbf{r})$. In particular, the ground state
solution of Eq.~(\ref{eq:EMTschr}) gives rise to a six-fold
degenerate donor ground state.

The symmetry group of the conduction band minima (and thus of the
Bloch functions $\varphi(\mathbf{r})$) is $C_{\infty v}$ in EMT,
which reduces to $C_{2v}$ in the silicon crystal. The envelope
wave functions $F(\mathbf{r})$ belong to $D_{\infty h}$. Their
products belong to the cross-section of both groups, which is
$C_{2v}$. For the $1s$-like ($m=0$) ground state function of
Eq.~\ref{eq:EMTschr} $F_\mu(\mathbf{r})$, such a product
transforms according to the $\Gamma_1$ representation of the
valley symmetry group $C_{2v}$. Because the donor is located at a
substitutional site of the tetrahedral silicon lattice, the total
wave function has $T_d$-symmetry. Using Frobenius' theorem
\cite{hamermesh62}, it can be shown that the $\Gamma_1$
representation of $C_{2v}$ induces the
$\Gamma_1+\Gamma_3+\Gamma_5$ representation \footnote{In
literature discussing donors in silicon, it is more common to
denote the single valued representations of $T_d$ by $A_1$, $A_2$,
$E$, $T_1$ and $T_2$, while for acceptors $\Gamma_i$ ($i=1\ldots
5$) is used. In this paper, we chose to use the
$\Gamma_i$-notation in all cases. Moreover, we use the
Sch\"onflies symbols to denote the crystallographic point groups
\cite{koster63}.} of $T_d$. This means that linear combinations of
the $F_\mu(\mathrm{r})$ can be found that have the correct
transformations properties under $T_d$. Using the notation
$\pmb{\alpha}=(\alpha_1,\ldots,\alpha_6)$ (as in
Eq.~\ref{eq:EMTgs}) the reduction to the $T_d$ representations is
carried out by
\begin{equation}
  \begin{array}{lll}
    \begin{array}{l}
      \pmb{\alpha}_g = \frac{1}{\sqrt{6}}(1,1,1,1,1,1) \\
    \end{array} & & \Gamma_1 \\[10pt]
    \begin{array}{l}
      \pmb{\alpha}_r = \frac{1}{\sqrt{12}}(-1,-1,-1,-1,2,2) \\
      \pmb{\alpha}_s = \frac{1}{2}(1,1,-1,-1,0,0) \\
    \end{array} &
    \left.\begin{array}{l}\ \\ \ \\ \end{array}\right\} &
      \Gamma_3 \\[20pt]
    \begin{array}{l}
      \pmb{\alpha}_x = \frac{1}{\sqrt{2}}(1,-1,0,0,0,0) \\
      \pmb{\alpha}_y = \frac{1}{\sqrt{2}}(0,0,1,-1,0,0) \\
      \pmb{\alpha}_z = \frac{1}{\sqrt{2}}(0,0,0,0,1,-1) \\
    \end{array} &
    \left.\begin{array}{l}\ \\ \ \\ \ \\ \end{array}\right\} &
    \Gamma_5 \\
  \end{array}
  \label{eq:alpha}
\end{equation}
Each of the vectors $\pmb{\alpha}$ defines a wave function $\Psi$
through Eq.~\ref{eq:EMTgs}. Here, the basis functions of the two-
and three dimensional representations have been chosen such that
$\Psi_r$ and $\Psi_s$ transform under $T_d$ as $3z^2-r^2$ and
$\sqrt{3}(x^2-y^2)$, respectively. Similarly, $\Psi_x$, $\Psi_y$
and $\Psi_z$ have been chosen such that they transform under $T_d$
as $x$, $y$ and $z$, respectively.

The potential term in the EMT-Schr\"odinger equation
(\ref{eq:EMTschr}) is a good approximation only for $r\gtrsim a$,
where $a$ is the lattice constant of silicon. For small $r$, the
charge of the nucleus is not screened by other electrons and it
will attract electrons much stronger than described by the
potential in Eq.~(\ref{eq:EMTschr}). Because the symmetry of the
potential is not affected, the states are still described by the
representations of $T_d$, but they are no longer degenerate. The
$\Gamma_1$ state $\Psi_g$ is the only one of the six ground state
wave functions that has non-zero electron density at the nucleus
($\mathbf{r}=0$). Therefore, it has a larger binding energy than
predicted by EMT and for most donors in silicon the $1s(\Gamma_1)$
state is the true ground state. This effect is generally called
`chemical splitting' (because the size of the effect depends on
the donor in question) or `valley-orbit splitting'. The remaining
states (especially the non-$s$ states) are quite well described by
EMT, because the electron density at the nucleus is negligible. As
an example, in case of phosphorus in silicon, the $1s(\Gamma_1)$
state (the ground state) has been measured to be located 45.29~meV
below the conduction band minimum \cite{ramdas81}, while the
EMT-prediction is $31.27$~meV \cite{faulkner69}.

\subsection{Symmetry of the donor ground state in an electric field}

After this brief review of established knowledge of silicon
donors, we return to the main subject of this paper. From purely
symmetry based considerations, we can find how the Hilbert
subspace spanned by the original six valley wave functions is
decomposed by the application of an electric field in a certain
direction. The impurities considered in this paper occupy
substitutional sites in the silicon lattice and their wave
functions transform according to representations of site symmetry
group $\bar{T}_d$. The symmetry group of a uniform electric field
$\bE$ is $C_{\infty v}$. When $\bE$ is applied in an arbitrary
direction in the silicon crystal, the symmetry group $\bar{T}_d$
of the Hamiltonian reduces to the trivial group $C_1$. Only when
the direction of the field is along one of the main
crystallographic directions of the crystal, the result is $C_{2v}$
for $\bE\parallel \langle 100\rangle$, $C_{3v}$ for $\bE\parallel
\langle 111\rangle$, and $C_s$ for $\bE\parallel \langle
110\rangle$. The reduction of symmetry can induce a splitting in
the original energy levels as shown in Table~\ref{tab:Esplit}. As
expected, the electric field does not remove degeneracy due to
time reversal symmetry and therefore all resulting levels are at
least two-fold degenerate.

\begin{table}
  \caption{Reduction of the site symmetry of an impurity in a uniform
  electric field in various directions and the resulting reduction of
  the irreducible representations \cite{koster63}.}
  \label{tab:Esplit}
  \centering
  \begin{tabular}{l|cccc}
  \hline
  \hline
  Direction & $\langle 100\rangle$ & $\langle 111\rangle$ &
    $\langle 110\rangle$ \\
  Group & $\bar{C}_{2v}$ & $\bar{C}_{3v}$ & $\bar{C}_s$\\
  \hline
  $\Gamma_1$ ($T_d$) & $\Gamma_1$ & $\Gamma_1$ & $\Gamma_1$ \\
  $\Gamma_2$ ($T_d$) & $\Gamma_3$ & $\Gamma_2$ & $\Gamma_2$ \\
  $\Gamma_3$ ($T_d$) & $\Gamma_1+\Gamma_3$ & $\Gamma_3$ &
    $\Gamma_1+\Gamma_2$  \\
  $\Gamma_4$ ($T_d$) & $\Gamma_2+\Gamma_3+\Gamma_4$ & $\Gamma_2+
    \Gamma_3$ & $\Gamma_1+2\Gamma_2$ \\
  $\Gamma_5$ ($T_d$) & $\Gamma_1+\Gamma_2+\Gamma_4$ & $\Gamma_1+
    \Gamma_3$ & $2\Gamma_1+\Gamma_2$ \\
  \hline
  $\Gamma_6$ ($\bar{T}_d$) & $\Gamma_5$ & $\Gamma_4$ &
    $\Gamma_{3+4}$ \\
  $\Gamma_7$ ($\bar{T}_d$) & $\Gamma_5$ & $\Gamma_4$ &
    $\Gamma_{3+4}$ \\
  $\Gamma_8$ ($\bar{T}_d$) & $2\Gamma_5$ & $\Gamma_4+\Gamma_{5+6}$ &
    $2\Gamma_{3+4}$ \\
  \hline
  \hline
  \end{tabular}
\end{table}

To make the connection to the valley wave functions
$F_\mu(\mathbf{r})\phi_\mu(\mathbf{r})$, we will now discuss the
symmetry of the $1s$ levels in an electric field from another
point of view. We start by looking at the individual valley wave
functions and subsequently derive which linear combinations form
appropriate donor wave functions (using the method of
Ref.~\onlinecite{ramdas63}). When a donor impurity in silicon is
situated in an electric field along the positive $z$-direction,
the valleys 5 and 6 keep their $C_{2v}$ symmetry, while the field
reduces the symmetry group of the other four valleys to $C_1$.
These four valleys are mixed by the elements of the site symmetry
group $C_{2v}$ and are therefore grouped together in the third
column of Table~\ref{tab:clasnovo}.

In case of a $1s$ state, the valley wave functions belong to the
$\Gamma_1$ representation of $C_{2v}$ (for valley 5 and 6) or
$C_1$ (for 1, 2, 3 and 4). This is found by reducing the even
$m=0$ representation of $D_{\infty h}$ to $C_{2v}$ and $C_1$,
respectively. By using Frobenius' theorem, it can be deduced that
these generate for the impurity wave function the representations
$\Gamma_1$ and $\Gamma_1 + \Gamma_2 + \Gamma_3 + \Gamma_4$ of
$C_{2v}$, respectively. This is also shown in
Table~\ref{tab:clasnovo}, together with the (set of) induced wave
function(s) spanning the subspace of that representation. In a
similar way, we obtained results for the electric field in the
other main crystallographic directions. They are also shown in the
table.

\begin{table}
  \caption{Considering the symmetry of the valley wave functions in an
  electric field, the symmetry of the total wave function they induce
  can be obtained. The results for the $1s$ level, without considering
  valley-orbit splitting, are shown in this table. The direction of $\bE$
  in the first column is denoted by the vectors defined in
  Fig.~\ref{fig:cube}(a). The fifth column lists
  the representations of the appropriate site symmetry group, given in the
  second column. The basis vectors
  are given in the notation of Eq.~(\ref{eq:alpha}).}
  \label{tab:clasnovo}
  \begin{tabular}{llllcl}
    \hline
    \hline
    \\[-10pt]
    Dir. & Site & Valley & Valley & $\Gamma$(site) & Basis \\
    $\bE$& sym. &       & sym. &                &       \\
    \hline
    $\mathbf{z}$ & $C_{2v}$&1, 2, 3, 4 & $C_1$ & $\Gamma_1$ & $(1,1,1,1,0,0)$ \\
                 &        &            &       & $\Gamma_2$ & $(1,-1,1,-1,0,0)$ \\
                 &        &            &       & $\Gamma_3$ & $(1,1,-1,-1,0,0)$ \\
                 &        &            &       & $\Gamma_4$ & $(1,-1,-1,1,0,0)$
                 \\[5pt]
                 &      & 5          & $C_{2v}$& $\Gamma_1$ & $(0,0,0,0,1,0)$
                 \\[5pt]
                 &      & 6          & $C_{2v}$& $\Gamma_1$ & $(0,0,0,0,0,1)$ \\
    \hline
    $\mathbf{w}$ & $C_{3v}$&1, 3, 5    & $C_s$ & $\Gamma_1$ & $(1,0,1,0,1,0)$ \\
                 &        &            &       & $\Gamma_3$ & $(\omega^2,0,\omega,0,1,0)$ \\
                 &        &            &       &            & $(\omega,0,\omega^2,0,1,0)$
                 \\[5pt]
                 &        & 2, 3, 6    & $C_s$ & $\Gamma_1$ & $(0,1,0,1,0,1)$ \\
                 &        &            &       & $\Gamma_3$ & $(0,\omega^2,0,\omega,0,1)$ \\
                 &        &            &       &            & $(0,\omega,0,\omega^2,0,1)$ \\
    \hline
    $\mathbf{v}$ & $C_s$  &1, 3       & $C_1$ &  $\Gamma_1$ & $(1,0,1,0,0,0)$ \\
                 &        &           &       &  $\Gamma_2$ & $(1,0,-1,0,0,0)$
                 \\[5pt]
                 &        &2, 4       & $C_1$ &  $\Gamma_1$ & $(0,1,0,1,0,0)$ \\
                 &        &           &       &  $\Gamma_2$ & $(0,1,0,-1,0,0)$
                 \\[5pt]
                 &        &5          & $C_s$ &  $\Gamma_1$ & $(0,0,0,0,1,0)$
                 \\[5pt]
                 &        &6          & $C_s$ &  $\Gamma_1$ & $(0,0,0,0,0,1)$ \\
    \hline
    \hline
  \end{tabular}
\end{table}

Due to the valley-orbit splitting (which has been ignored so far)
the three irreducible components of the donor ground state are
already energetically separated at zero field. Therefore, the
basis vectors have to be chosen in such a way that they agree with
the zero-field energy splitting of the $\Gamma_1$, $\Gamma_3$ and
$\Gamma_5$ levels of $T_d$ \footnote{It is known from group theory
that the reduction of a representation containing more than one
instance of the same irreducible representation is not uniquely
determined.}. The result for various directions of the electric
field is shown in Table~\ref{tab:clasvo}.

\begin{table*}
  \caption{Reduction of the $1s$ donor energy levels in an electric
  field. The basis vectors belonging to these states are given (in the
  notation of Eq.~(\ref{eq:alpha})) in the limit $\mcE\rightarrow 0$
  ($\omega=e^{2\pi i/3}$).
  The eigenvalues (up to second order in $\mcE$) are the result of the
  perturbation calculation described in the text.}
  \label{tab:clasvo}
  \begin{tabular}{lccll}
    \hline
    \hline
    Field direction & $\mcE=0$ & $\mcE\neq 0$  & Basis vector(s) & Eigenvalue\\[5pt]
    \hline
    $\mathbf{z}$ & $\Gamma_1(T_d)$ & $\Gamma_1(C_{2v})$ & $(1,1,1,1,1,1)/\sqrt{6}$ &
      $E_1-\frac{|p_{15}|^2}{E_5-E_1}\mcE^2$ \\[10pt]
    & $\Gamma_3(T_d)$ & $\Gamma_1(C_{2v})$ & $(1,1,1,1,-2,-2)/\sqrt{12}$ &
      $E_3 +\frac{|2p_{35}|^2}{E_3-E_1}\mcE^2$ \\
    &            & $\Gamma_3(C_{2v})$ & $(1,1,-1,1,0,0)/2$ & $E_3$\\[10pt]
    & $\Gamma_5(T_d)$ & $\Gamma_1(C_{2v})$ & $(0,0,0,0,1,-1)/\sqrt{2}$ &
      $E_5+(\frac{|p_{15}|^2}{E_5-E_1}+\frac{|2p_{35}|^2}{E_3-E_1})\mcE^2$\\
    &            & $\Gamma_2(C_{2v})$ & $(1,-1,1,-1,0,0)/\sqrt{2}$ & $E_5+|p_5|\mcE$\\
    &            & $\Gamma_4(C_{2v})$ & $(1,-1,-1,1,0,0)/\sqrt{2}$ & $E_5-|p_5|\mcE$\\[5pt]
    \hline
    $\mathbf{w}$ & $\Gamma_1(T_d)$ & $\Gamma_1(C_{3v})$ & $(1,1,1,1,1,1)/\sqrt{6}$ &
      $E_1-\frac{|p_{15}|^2}{E_5-E_1}\mcE^2$\\[10pt]
    & $\Gamma_3(T_d)$ & $\Gamma_3(C_{3v})$ & $(\omega^2,\omega^2,\omega,\omega,1,1)/\sqrt{6}$ &
      $E_3+\frac{2|p_{35}|^2}{E_3-E_5}\mcE^2$ \\
    &            &            & $(\omega,\omega,\omega^2,\omega^2,1,1)/\sqrt{6}$ &\\[10pt]
    & $\Gamma_5(T_d)$ & $\Gamma_1(C_{3v})$ & $(1,-1,1,-1,1,-1)/\sqrt{6}$ &
      $E_5\pm\frac{2}{3}\sqrt{3}|p_5|\mcE+
        (\frac{|p_{15}|^2}{E_5-E_1}-\frac{4|p_{35}|^2}{E_3-E_5})\mcE^2$\\
    &            & $\Gamma_3(C_{3v})$ & $(\omega^2,-\omega^2,\omega,-\omega,1,-1)/\sqrt{6}$ &
      $E_5\mp\frac{1}{3}\sqrt{3}|p_5|\mcE$\\
    &            &            & $(\omega,-\omega,\omega^2,-\omega^2,1,-1)/\sqrt{6}$ & \\[5pt]
    \hline
    $\mathbf{v}$ & $\Gamma_1(T_d)$ & $\Gamma_1(C_s)$ & $(1,1,1,1,1,1)/\sqrt{6}$ &
      $E_1-\frac{|p_{15}|^2}{E_5-E_1}\mcE^2$\\[10pt]
    & $\Gamma_3(T_d)$ & $\Gamma_1(C_s)$ & $(1,1,1,1,-2,-2)/\sqrt{12}$ &
      $E_3 +\frac{|p_{35}|^2}{E_3-E_1}\mcE^2$\\
    &            & $\Gamma_2(C_s)$ & $(1,1,-1,1,0,0)/2$ &
      $E_3+\frac{3|p_{35}|^2}{E_3-E_1}\mcE^2$\\[10pt]
    & $\Gamma_5(T_d)$ & $\Gamma_1(C_s)$ & $(0,0,0,0,1,-1)/\sqrt{2}$ &
      $E_5+|p_5|\mcE-\frac{1}{2}(\frac{|p_{35}|^2}{E_3-E_1}-\frac{|p_{15}|^2}{E_5-E_1})\mcE^2$ \\
    &            & $\Gamma_1(C_s)$ & $(1,-1,1,-1,0,0)/\sqrt{2}$ &
      $E_5-|p_5|\mcE-\frac{1}{2}(\frac{|p_{35}|^2}{E_3-E_1}-\frac{|p_{15}|^2}{E_5-E_1})\mcE^2$ \\
    &            & $\Gamma_2(C_s)$ & $(1,-1,-1,1,0,0)/\sqrt{2}$ &
      $E_5-\frac{3|p_{35}|^2}{E_3-E_1}\mcE^2$ \\[5pt]
    \hline
    \hline
  \end{tabular}
\end{table*}

\subsection{Shift and splitting in an electric field}

Now, we will derive the shift and splitting of the lowest donor
levels in an electric field from a perturbation calculation.
Results for other levels can be derived using the same method,
although (because the level spacing is smaller for higher levels)
the range of fields where the perturbation calculation is valid is
much smaller.

Although the six-fold degeneracy of the $1s$-levels is lifted by
the valley orbit interaction, the complete manifold is relatively
well-separated from the higher levels (the separation of the
highest $1s(\Gamma_3)$ level to closest exited level ($2p_0$) is
roughly twice as large as the separation between the
$1s(\Gamma_1)$ and $1s(\Gamma_3)$ levels). Therefore, we consider
the $1s$-manifold as a whole in a single perturbation calculation,
taking only the coupling among the $1s$ levels themselves into
account.

The electric field couples to the (induced) dipole moment
$\mathbf{D}= e\mathbf{r}$ of the impurity state and gives rise to
an additional term in its Hamiltonian $-\bE\cdot\mathbf{D}$,
reflecting the energy associated with the dipole in the field. By
making use of the Wigner-Eckart orthogonality theorem from group
theory \cite{tsukerblat94}, it is possible to find the vanishing
matrix elements as well as the dependencies between the
non-vanishing matrix elements, as they follow from the symmetry of
the system. The $1s$ sub-matrix $[\Hmlt]$ of the total Stark
Hamiltonian $\Hmlt = \Hmlt_0 + \bE\cdot\mathbf{D}$ is given by
$$
  \left(\begin{array}{c|cc|ccc}
    E_1 & 0 & 0 & p_{15}\mcE_x & p_{15}\mcE_y & p_{15}\mcE_z \\
    \hline
    0 & E_3 & 0 & -p_{35}\mcE_x & -p_{35}\mcE_y &
      2p_{35}\mcE_z \\
    0 & 0 & E_3 & p_{35}\sqrt{3}\mcE_x & -p_{35}\sqrt{3}\mcE_y & 0 \\
    \hline
    \bar{p}_{15}\mcE_x & -\bar{p}_{35}\mcE_x & \bar{p}_{35}\sqrt{3}\mcE_x &
      E_5 & p_5\mcE_z & p_5\mcE_y \\
    \bar{p}_{15}\mcE_y & -\bar{p}_{35}\mcE_y & \bar{p}_{35}\sqrt{3}\mcE_y &
      \bar{p}_5\mcE_z & E_5 & p_5\mcE_x \\
    \bar{p}_{15}\mcE_z & 2\bar{p}_{35}\mcE_z & 0 & \bar{p}_5\mcE_y &
      \bar{p}_5\mcE_x & E_5 \\
  \end{array}\right).
$$
The elements of this matrix are given by $[\Hmlt]_{ij}= \langle
\varphi_i|\Hmlt|\varphi_j\ \rangle$, where the wave functions
$\varphi_i$ are taken from the basis $(\Psi_g, \Psi_r, \Psi_s,
\Psi_x, \Psi_y, \Psi_z)$) as defined before. The energies $E_1$,
$E_3$ and $E_5$ are the eigenvalues of the unperturbed Hamiltonian
$\Hmlt_0$, that is the zero-field energies of the $1s(\Gamma_1)$,
$1s(\Gamma_3)$ and $1s(\Gamma_5)$ level, respectively. For
phosphorous in silicon, the values are $E_1=-45.59$~meV,
$E_3=-32.58$~meV, and $E_5=-33.89$~meV with respect to the
conduction band edge \cite{ramdas81}. The parameters $p_{15}$,
$p_{35}$ and $p_5$ describe the coupling between the $1s$-levels.
As can be seen, these are the only three independent parameters
describing the coupling between the levels. They can be expressed
in terms of integrals over products of wave functions, e.g.
$p_{15}=e\langle \Psi_g|x| \Psi_x \rangle$ and $p_{5}=e\langle
\Psi_y|x| \Psi_z \rangle$.

\begin{figure}
  \centering
  \includegraphics[width=8.6cm]{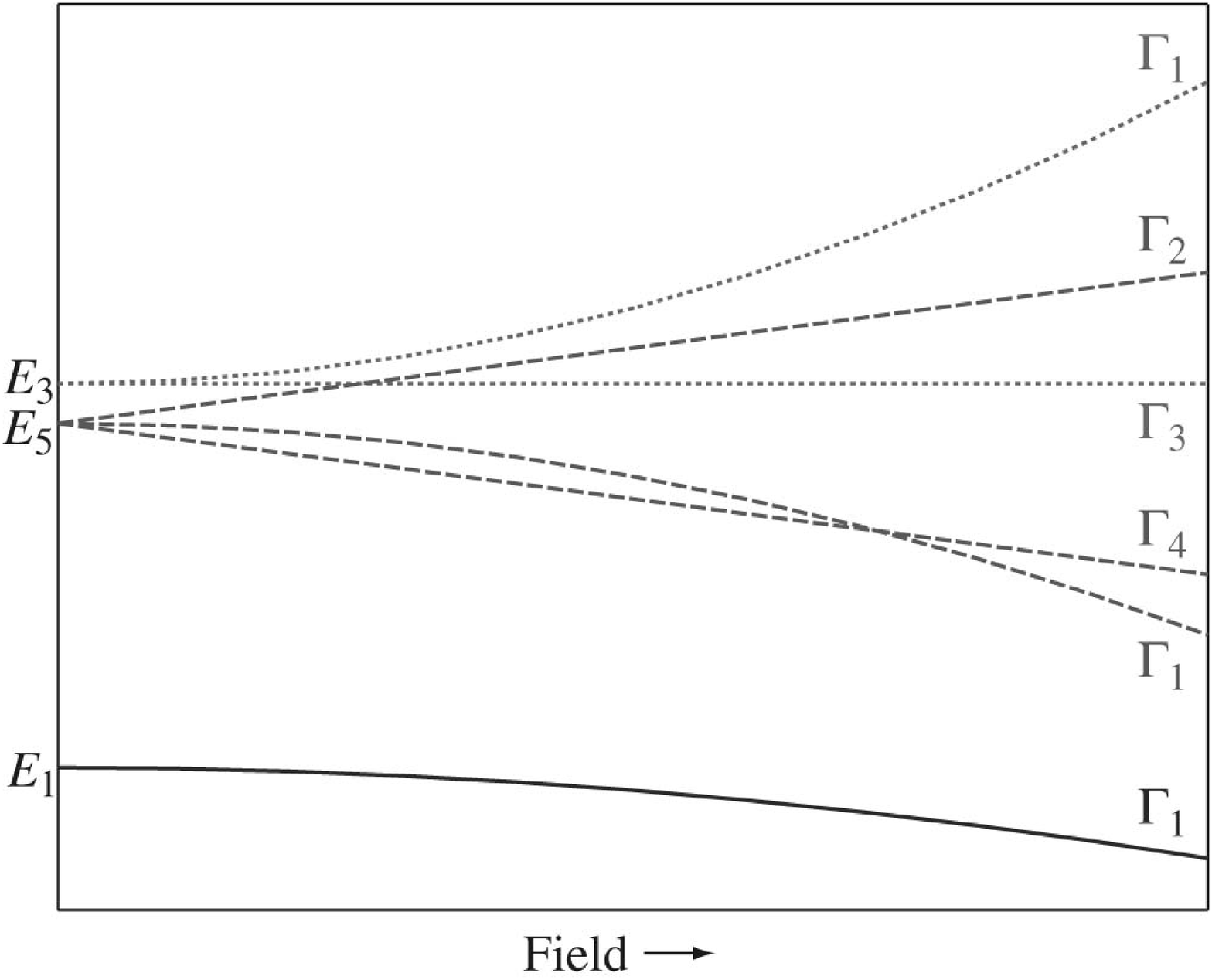}
  \caption{Schematic plot of the $1s$ energy levels as a function of
  the electric field $\mcE$. The values of the parameters $p_5$, $p_{15}$,
  and $p_{35}$ have been chosen such that the plot clearly illustrates the
  qualitative features of the Stark effect in the energy levels.}
  \label{fig:split}
\end{figure}

Perturbation theory is invoked by calculating the eigenvalues and
eigenvectors of this $6\times 6$ matrix up to second order in
$\mcE$. This yields the $1s$ energy levels and wave functions as a
function of electric field for $\bE$ along the three main
crystallographic directions. The energy levels are presented in
the last column of Table~\ref{tab:clasvo}. From
Table~\ref{tab:clasvo} it can be seen that the $1s(\Gamma_1)$
ground state experiences an isotropic quadratic shift downwards
\footnote{In general, the shift of a $\Gamma_1$ level cannot have
any dependence on the direction of the field.}, while for the
other levels the behavior depends on the direction of the electric
field. In Figure~\ref{fig:split} the results for $\bE\parallel
\langle 100\rangle$ are plotted schematically.

The corresponding eigenvectors were also obtained from this
calculation. In the limit $\mcE\rightarrow 0$ they coincide with
the vectors given in Table~\ref{tab:clasvo}, allowing to label
each eigenvalue with the correct representation. These results are
directly applicable in the prediction of allowed optical
transitions between the various levels.

We discuss the behavior of the three $1s$ states in  in some more
detail. The normalized eigenfunctions in an electric field
parallel to $\mathbf{z}$ (again up to second order in $\mcE$)
corresponding to the eigenvalues already given in
Table~\ref{tab:clasvo} are
\begin{equation}
  \begin{array}{l}
    \Phi_g=(1-\textstyle\frac{1}{2}|\beta|^2\mcE^2)\Psi_g+
      \beta''\mcE^2\cdot\Psi_r - \bar{\beta}\mcE\cdot\Psi_z \\[10pt]
    \Phi_r=-\bar{\beta}''\mcE^2\cdot\Psi_g+
      (1-\textstyle\frac{1}{2}|\beta'|^2\mcE^2)\Psi_r + \bar{\beta}'\mcE\cdot\Psi_z \\
    \Phi_s=\Psi_s \\[10pt]
    \Phi_x=\textstyle\frac{1}{2}\sqrt{2}(\Psi_x+\Psi_y) \\
    \Phi_y=\textstyle\frac{1}{2}\sqrt{2}(\Psi_x-\Psi_y) \\
    \Phi_z=\beta\mcE\cdot\Psi_g-\beta'\mcE\cdot\Psi_r+\\
      \multicolumn{1}{r}{(1-\textstyle\frac{1}{2}(|\beta|^2+|\beta'|^2)\mcE^2)\Psi_z} \\
  \end{array}
  \label{eq:eigfun}
\end{equation}
where
\begin{equation*}
  \beta=\frac{p_{15}}{E_5-E_1},\ \
  \beta'=\frac{2p_{35}}{E_3-E_5}, \ \
  \beta''=\bar{\beta}\frac{2p_{35}}{E_3-E_1}.
\end{equation*}

The initial zero field wave function $\Psi_g$ has the highest
spacial symmetry possible in a tetrahedral lattice. To get more
insight in the contribution of the six valleys as a function of
the applied field, we can write the perturbed ground state wave
function $\Phi_g$ in the notation of Eq.~(\ref{eq:alpha}) as
\begin{multline*}
  (1,1,1,1,1,1) + (0,0,0,0,-\gamma'',\gamma'')\mcE + \\
  (-\gamma,-\gamma,-\gamma,-\gamma,-\gamma',-\gamma')\mcE^2,
\end{multline*}
where
$$
  \gamma=\frac{1}{2}(|\beta|^2+\beta''\sqrt{2}), \ \
  \gamma'=\frac{1}{2}(|\beta|^2-2\beta''\sqrt{2}), \ \
  \gamma''=\bar{\beta}\sqrt{3}
$$
and an overall factor $1/\sqrt{6}$ was omitted. From these
expressions, we see that the contribution of the valley in the
$-\mathbf{z}$ direction increases linearly with the field, while
contribution of the opposite valley decreases linearly with the
field. This reflects the field-induced dipole moment of the ground
state.

The results of this calculation could be made quantitative if the
values of the parameters $p_5$, $p_{15}$, and $p_{35}$ were known.
This can be done by evaluating the integrals defining these
parameters and using e.g.\ the EMT wave functions from
Eq.~(\ref{eq:EMTgs}). However, due to the strongly oscillating
integrants, this is numerically a non-trivial task. Furthermore,
the EMT wave functions have a higher symmetry than the lattice,
and the value for $p_5$ obtained in this way is always zero. An
estimate for $p_5$ can only be obtained using more sophisticated
approximations for the wave functions. More importantly, the
applicability of such results is limited, especially for the $1s$
state, as the effects of valley-orbit interaction are not included
in the EMT wave functions.

It is important to note that the energies in
Table~\ref{tab:clasvo} and the eigenstates in
Eq.~(\ref{eq:eigfun}) are based on symmetry properties only and
not on the explicit form of the EMT wave functions. Therefore,
these results remain valid, even if valley-orbit interaction and
central cell corrections are fully included. Such modifications
would only influence the values of the parameters $p_5$, $p_{15}$,
and $p_{35}$.

\section{Acceptors}

\label{sec:acceptors}

Acceptor wave functions can be equally well used for ASE as
donors. Recent experiments showing that the coherence time of
spins of bound holes is more than 1~ms \cite{golding03}, even
justify the prospective use of acceptor wave functions as qubits.
We therefore also briefly outline the properties of silicon
acceptors in an electric field, taking the silicon valence band
structure into account. The initially threefold degenerate valence
band maximum is split by spin orbit interaction, which causes one
of the bands to shift downwards by $\sim 43$~meV \cite{ramdas81}.
Due to the spin-orbit interaction, spin is not a good quantum
number anymore and the bands must be characterized by the total
angular momentum, which is $\frac{3}{2}$ for the upper two bands.
Due to the half-valued angular momentum, the Bloch wave function
at the valence band maximum transforms according to one of the
double valued representations of $\bar{T}_d$, namely $\Gamma_8$.
As a result, the total impurity wave functions transform according
to representations of the same group. The ground state wave
function, as well as the first few excited levels belong to the
$\Gamma_8$ representation and they are all four-fold degenerate
(including spin).

\subsection{Linear Stark effect}

To derive the small-field splitting of acceptors in silicon in an
electric field, we use degenerate perturbation theory for each
level individually. To that end, the Hamiltonian sub-matrix
$\langle \varphi_i |\Hmlt| \varphi_j \rangle$ of the level under
consideration must be calculated and diagonalized, where the
$\varphi_i$ form a suitable basis for the subspace of that
particular level.

As mentioned before, the components of the electric dipole
operator $e\mathbf{r}$ transform according to the rows of the
$\Gamma_5$ representation of $\bar{T}_d$. Because the
anti-symmetrized direct products $\{\Gamma_6 \times \Gamma_6\} =
\{\Gamma_7 \times \Gamma_7\} = \Gamma_1$ do not contain
$\Gamma_5$, the first order Stark matrix vanishes for levels with
$\Gamma_6$ or $\Gamma_7$ symmetry. Hence, such levels do not
experience a linear Stark effect. On the other hand, $\{\Gamma_8
\times \Gamma_8\}= \Gamma_1+ \Gamma_3+ \Gamma_5$ does contain
$\Gamma_5$, so that a linear Stark effect is possible for a
$\Gamma_8$ level \footnote{Note that a substitutional site in
silicon has no inversion symmetry and therefore no definite
parity. This is essential for the occurrence of a linear Stark
effect in an isolated level.}.

The effective linear Stark Hamiltonian \footnote{In contrast to
our treatment of donors, we will use the technique of effective
Hamiltonians to derive the matrices for acceptor levels.} for a
$\Gamma_8$ level is given by \cite{bir63b}
$$
  [\Hmlt]_8^\mathrm{lin}=
    \frac{2}{\sqrt{3}}p_8\big(\mcE_x\{J_y,J_z\}+\mcE_y\{J_z,J_x\}+
      \mcE_z\{J_x,J_y\}\big),
$$
where the parameter $p_8$ is related to the effective dipole
moment of such a state. The $J_i$ ($i=x,y,z$) are matrices of the
components of the angular momentum operator with respect to some
convenient basis and $\{A,B\}=\frac{1}{2}(AB+BA)$ is the
anti-commutator. The eigenvalues of this matrix are given by
$$
  E_8 \pm |p_8|\mcE,
$$
where both eigenvalues occur twice. This is a symmetric splitting
of the level, which is independent of the direction of $\bE$. Note
that $p_8$ vanishes within EMT, similar to $p_5$ before. Estimates
of $p_8$ obtained in literature range from $10^{-2}$~D
\cite{bir63b} to 0.26~D \cite{kopf92}.

\subsection{Quadratic Stark effect}

Because $\{\Gamma_6\times\Gamma_6\}= \{\Gamma_7\times\Gamma_7\}=
\Gamma_1$, the quadratic effective Stark-Hamiltonian for a
$\Gamma_6$ and $\Gamma_7$ level is simply given by
$$
  \Hmlt_\mathrm{eff,quad}=a_i \mcE^2 \hat{I},
$$
where $\hat{I}$ is the identity matrix and the $a_i$ ($i=6,7$) are
phenomenological parameters, that can be expressed in terms of
integrals over wave functions. It follows that the $\Gamma_6$ and
$\Gamma_7$ levels experience an isotropic quadratic shift
$$
  E_i + a_i\mcE^2,
$$
where $E_i$ is the unperturbed energy of a $\Gamma_i$ level. The
two-fold degeneracy due to time reversal symmetry is obviously not
removed by the electric field.

The quadratic part of the effective Hamiltonian for a $\Gamma_8$
level, such as the ground state, is given by \cite{bir63b}
$$
  \begin{array}{l}
  [\Hmlt]_8^\mathrm{quad}=a_8 \mcE^2 \hat{I} +
    b_8\big[J_x^2\mcE_x^2+J_y^2\mcE_y^2+J_z^2\mcE_z^2-
    \frac{1}{3}\mathbf{J}^2\big]\\
    \ \ +\frac{2}{\sqrt{3}}c_8\big[\{J_x,J_y\}\mcE_x\mcE_y+
    \{J_y,J_z\}\mcE_y\mcE_z+\{J_z,J_x\}\mcE_z\mcE_x\big],
  \end{array}
$$
where $a_8$, $b_8$ and $c_8$ are again phenomenological
parameters. The total Hamiltonian has two distinct eigenvalues
\begin{multline}
  a_8\mcE^2\pm\big[p^2\mcE^2+b_8^2\mcE^4 \\
  +(c_8^2-3b_8^2)(\mcE_y^2\mcE_z^2+\mcE_x^2\mcE_z^2+\mcE_x^2\mcE_y^2) \\
  +6p_8 c_8\mcE_x\mcE_y\mcE_z\big]^{1/2},\\
\end{multline}
each of which is still doubly degenerate (due to time reversal
symmetry) \footnote{Note that there is a mistake in the
corresponding expression in Ref.~\onlinecite{bir63b}, where the
last term between the square brackets is missing.}. For
$\bE\parallel \langle 100\rangle$ this expression reduces to (up
to second order in $\mcE$)
$$
  E_8 \pm |p_8|\mcE+a_8\mcE^2.
$$
For $\bE\parallel \langle 111\rangle$ we find
$$
  E_8 \pm |p_8|\mcE+(a_8\pm\frac{1}{3}\sqrt{3}c_8)\mcE^2
$$
and for $\bE\parallel \langle 110\rangle$ we have
$$
  E_8 \pm |p_8|\mcE+a_8\mcE^2.
$$
The results for $\bE\parallel \langle 100\rangle$ and for
$\bE\parallel \langle 110\rangle$ are the same in this
approximation, but different in third order.

Obviously, the wave functions of donors and acceptors are very
different and this is reflected in their respective electric field
behavior. The donor ground state undergoes an isotropic quadratic
shift. The acceptor ground state has an isotropic linear
splitting, superposed on an anisotropic quadratic shift.

In the spectroscopic measurements of boron acceptors in silicon
\cite{white67} (mentioned in the introduction), the observed
$\Gamma_8$-levels indeed show a quadratic shift. However, the
expected level-splitting was not observed, most likely due to
limited resolution.

\section{Large electric fields in SHM}

\label{sec:H}

In this section, we will calculate energy levels of an impurity in
a semiconductor as a function of electric field in the range from
zero to $\sim 5$~MV/m. This is done within the scaled hydrogen
model, where the band structure of the semiconductor is accounted
for by a single effective mass and the dielectric constant only.

For this calculation it is convenient to express all quantities in
so-called effective atomic units. For instance, energies are
expressed in units of twice the effective ionization energy and
length in units of the effective Bohr-radius. Conversion of units
of relevant quantities for both vacuum and silicon are given in
Table~\ref{tab:au}.

In the past, several algorithms have been described in literature
to calculate electric field dependence of the energy levels of the
hydrogen atom. However, very little results in the range of
interest for ASE (fields up to $\sim 0.1$~a.u. \cite{kane00}) have
been published. Therefore, we found it important to fill this gap
by fully presenting the results of our calculation. For this
purpose, we used the slightly adapted version of a variational
algorithm that not only yields the energy levels, but also their
lifetimes \cite{alvarez91}.

\begin{table}
  \caption{Atomic units for some relevant physical quantities in vacuum
  and silicon. For silicon the values $\es=11.4$ and $m^*=0.26$ (appropriate
  for electrons) were taken.}
  \label{tab:au}
  \centering
  \begin{tabular}{l|lrr}
    \hline
    \hline
    Quantity & Unit & Value in vacuum & Value in Si \\
    \hline
    Energy & 2Ry & 27.2~eV & 54~meV \\
    Length & $a_0$ & 0.053~nm & 2.3~nm \\
    Electric field & $2\mathrm{Ry}/ea_0$ & 510~GV/m & 24~MV/m \\
%    Dipole moment & $ea_0$
%      & $8.5\cdot 10^{-30}$~Cm & $3.7\cdot 10^{-28}$~Cm\\
    Time & $\hbar/2\mathrm{Ry} $
      & $2.4\cdot 10^{-17}$~s & $1.2\cdot 10^{-14}$~s \\
    \hline
    \hline
  \end{tabular}
\end{table}

For completeness, we will very briefly outline the main features
of this method. The hydrogen Schr\"odinger equation (including the
electric field) in parabolic coordinates can be separated, which
allows for high numerical accuracy without too much computational
effort. In order to be able to find the energy positions of the
resonances as well as their lifetimes, the complex scaling method
was applied \cite{moiseyev98}. Then, for each coordinate the
Hamiltonian (including electric field) is expanded with respect to
a truncated basis of unperturbed wave functions. This can be done
analytically. Finally, the energy levels and lifetimes are
obtained by tracking (separately for each level) the eigenvalues
of this matrix from zero field in small steps to larger fields.

\begin{figure}
  \centering
  \includegraphics[width=8.6cm]{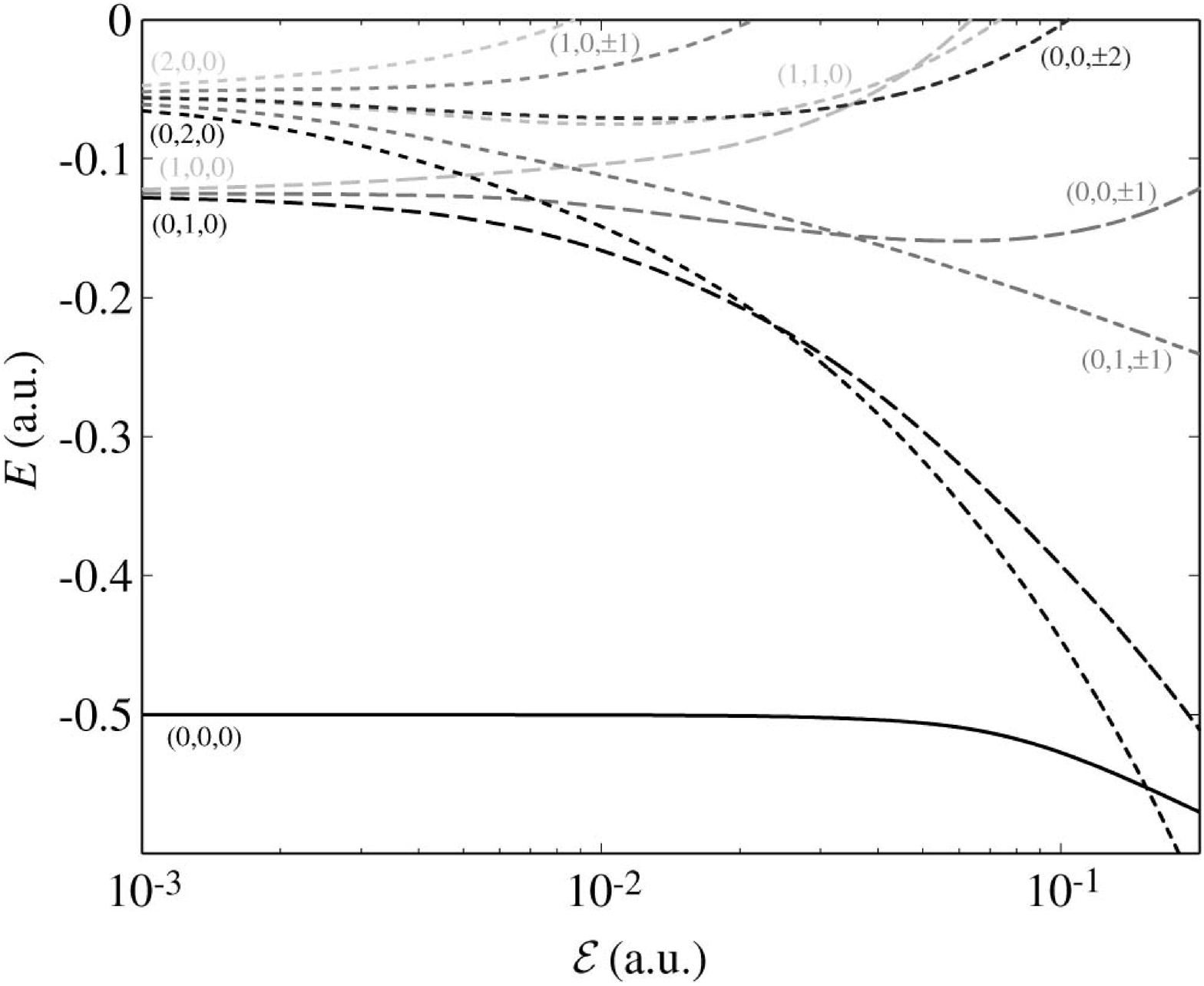}
  \caption{Evolution of the lowest lying energy levels ($n=1,2,3$)
  of a hydrogen-like system versus electric field $\mcE$. For conversion
  of a.u.\ to conventional units, see Table~\ref{tab:au}.}
  \label{fig:eigE}
\end{figure}

By using the method described above, we calculated the energies of
all states with $n=1,2,3$ for $0\leq \mcE\leq 0.2$\,a.u. The
results for the energy levels are depicted in Fig.~\ref{fig:eigE}.
The levels are labelled by parabolic quantum numbers
\cite{bethe57} $(n_1,n_2,m)$, which are more suitable for hydrogen
in an electric field than the more common spherical quantum
numbers $(n,l,m)$. The magnetic quantum number $m$ has the same
meaning in both representations. The main quantum number $n$ is
related to the parabolic quantum numbers by $n=n_1+n_2+|m|+1$. The
electric field lifts all degeneracies except for spin and
$(n_1,n_2,\pm m)$. So (including spin) there are both two-fold
degenerate levels ($m=0$) and fourfold degenerate levels ($m\neq
0$).

Figure~\ref{fig:eigE} shows that the ground state ($n=1$) exhibits
a small second order shift downwards. The $n=2$-level splits into
three levels. Two of them are (for small $\mcE$) linearly shifting
upwards and downwards. The middle one has no first order shift,
consistent with the well known results from perturbation theory
\cite{bethe57}. Finally, the ninefold degenerate $n=3$-level can
be seen to split into six levels. As expected, the effect of the
electric field on higher levels is larger, due to their larger
spatial extent. At large values of the field, several levels cross
each other \footnote{Levels belonging to the same representation
of the spatial symmetry group $C_{\infty v}$ can be seen to cross
each other in Fig.~\ref{fig:eigE}. This is however no violation of
the non-crossing rule, since for this specific problem there
exists an additional constant of motion that is associated with
the separability of the Hamiltonian \cite{helfrich72}.} and some
of them show non-monotonous behavior.

The few results of calculations that can be found in literature
(obtained by different methods) and overlap with our results are
in very good agreement, both for the ground state \cite{ivanov97}
and for the first excited state ($m=1$) \cite{fernandez96}.

\begin{figure}
  \centering
  \includegraphics[width=8.6cm]{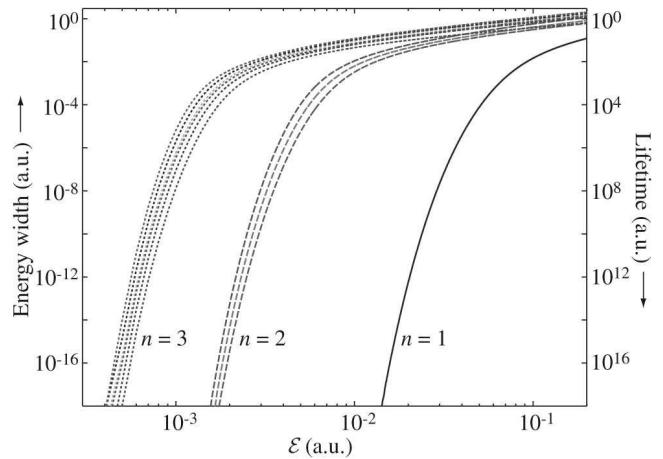}
  \caption{Energy width and lifetime of the lowest lying energy
  levels of hydrogen-like systems ($n=1,2,3$) versus electric field
  $\mcE$. For conversion of a.u.\ to conventional units, see
  Table~\ref{tab:au}.}
  \label{fig:eigdE}
\end{figure}

The method we used for our calculations can not only be extended
to very large fields, but it also has the advantage of yielding
the width of the energy levels. The increasing energy width of the
hydrogen-like levels in an electric field is the results of the
ability of the field to ionize the atom. The finite probability
for the carrier to tunnel out of the nucleus' potential well leads
to a finite lifetime \footnote{This lifetime is solely due to the
possibility of ionization and is unrelated to (radiative or
non-radiative) transitions from an excited level to a lower
state.} of the level. In Fig.~\ref{fig:eigdE}, the evolution of
the width of several hydrogen energy levels is depicted.
Obviously, the width of all levels is zero at zero field, which is
equivalent to an infinitely long lifetime. For any nonzero $\mcE$,
the lifetimes have a finite value, that decreases monotonously
with the field. The stronger the binding energy of a level at zero
field, the faster the lifetime decreases when the field increases.

\begin{figure}
  \centering
  \includegraphics[width=8.6cm]{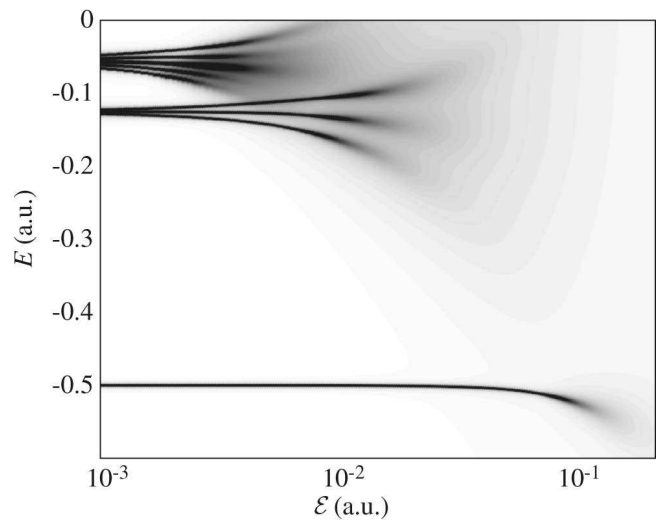}
  \caption{Map of the energy levels from Figure~\ref{fig:eigE}, converted
  to Lorentzians using the data of Figure~\ref{fig:eigdE}. For conversion
  of a.u.\ to conventional units, see Table~\ref{tab:au}.}
  \label{fig:starkmap}
\end{figure}

In Figure~\ref{fig:starkmap}, the results of Figure~\ref{fig:eigE}
and~\ref{fig:eigdE} are combined into one `intensity map', where
the levels are displayed as normalized Lorentzian line shapes, the
width of which is taken from Fig.~\ref{fig:eigdE}. The figure
shows clearly that for the realistic electric field
$\mcE=0.04$~a.u. (about 1~MV/m; see Table~\ref{tab:au}) the energy
width of all levels except the ground state is already larger than
or comparable to their binding energy. The ground state lifetime
is only 10~ns at that field. We  also note that for our purpose it
is not very useful to extend the calculation to higher fields, as
already at $\mcE=0.2$~a.u. all levels are very much broadened and
strongly overlapping. Although in case of hydrogen atoms in vacuum
such large field (0.2~a.u.\ $\sim 100$~GV/m) are only realized in
astronomy, in semiconductors they can be easily achieved under
laboratory conditions (0.2~a.u.\ $\sim 5$~MV/m).

Though the SHM oversimplifies the bandstructure, it is in our
opinion particularly useful to estimate lifetimes.
Fig.~\ref{fig:eigdE} shows that the lifetimes are primarily a
function of the zero-field binding energies. Assuming this is
still true when the silicon bandstructure is included,
interpolation of the results can be expected to provide a good
first order approximation of the level's true lifetime. For
example, the $n=1$ value in Fig.~\ref{fig:eigdE} underestimates
the phosphorous donor ground state lifetime, because it is
stronger bound than assumed in EMT.

When the electric field is generated by a small local gate, this
gate is usually separated from the semiconductor by a potential
barrier that is sufficiently high to prevent tunneling. If the
distance of the dopant atom to the barrier is not too small,
ionization of the dopant atom can still occur in large fields (and
the lifetimes discussed before still apply). However, the charge
carrier will not be `lost', but transferred to the potential well
created by the biased gate \cite{smit03}.

\section{Discussion and conclusion}

\label{sec:dis}

In the preceding sections, we have used two distinct approaches to
study the behavior of impurity wave functions in an electric
field. The first includes details of the bandstructure, but is
only valid for small fields and is somewhat qualitative. From this
symmetry-based analysis, we derived the energy level shift and
splitting for donors and acceptors in small electric fields, as
well as the modification of the donor wave function. Furthermore,
the symmetry classification of the resulting levels provides for
straightforward prediction of allowed optical transitions.

The second approach, the scaled hydrogen model, is fully
quantitative and applicable up to large fields, but neglects most
features of the silicon bandstructure. Still, the SHM offers a
manageable and valuable way to describe important phenomena in
atomic scale electronics. We presented the energy width and
lifetime of the impurity levels in large electric fields,
calculated within this framework.

It is possible to combine the two approaches and treat
Eq.~(\ref{eq:EMTschr}) in a way similar to that presented in
Section~\ref{sec:H}. Though this is in principle straightforward,
the reduced symmetry and lack of separability will make this
approach numerically very involved. Furthermore, it is important
to note that the direction of the electric field with respect to
the valley axis is not the same for all valleys. As an example,
for $\mcE\parallel \mathbf{z}$ the energy levels of $F_5$ and
$F_6$ are affected in a different way than those of the other four
$F_\mu$. If the solutions for the various valley wave functions
are known, they can be combined into impurity wave functions using
the data in Table~\ref{tab:clasnovo}.

Though potentially interesting, such an effort is not likely to
yield a good description of the dopant's wave function at high
electric fields, despite the tremendous increase of necessary
computational power. The reason is the omission of valley orbit
interaction, which not only affects the ground state, but also the
coupling to excited states. Especially for large fields, the
coupling influences the properties \emph{all} energy levels. It
has been shown that inter-valley coupling accounts for the
splitting of the $1s$ state for P in Si \cite{balderischi70}.
Inclusion of this effect appears to be a minimum requirement for
obtaining accurate quantitative results valid at large fields.

Recently, calculations of a silicon donor in an electric field in
the tight binding approach have been presented \cite{martins03}.
This approach seems to be a useful alternative to calculations
based on effective mass theory. Given the fact that this method
inherently includes the band structure of the semiconductor host,
it is striking how similar the results are to calculations based
on the SHM \cite{smit03}. This underlines the power of the SHM in
this type of calculations.

In summary, we have calculated the Stark effect of impurities in
silicon in two different approaches. Moreover, we discussed the
results and the computation methods used in the context of atomic
scale electronics and quantum computation.

\begin{acknowledgments}
One of us, S.R., wishes to acknowledge the Royal Netherlands
Academy of Arts and Sciences for financial support.
\end{acknowledgments}

\appendix

\section{Character tables}

In this appendix, we give for completeness the character tables of
various symmetry groups, relevant for this paper.
Table~\ref{tab:charTd} refers to the lattice symmetry group
$\bar{T}_d$. Depending on the direction of the electric field, it
reduces to one of the groups $C_{2v}$, $C_{3v}$ or $C_s$, the
character tables of which are given in Table~\ref{tab:charC2v}.
Finally, the table of the continuous groups $D_{\infty h}$ and
$C_{\infty v}$ are given in Table~\ref{tab:charDih}.

\begin{table}
  \caption{Character table for the double group $\bar{T}_d$.}
  \label{tab:charTd}
  \centering
  \begin{tabular}{l|cccccccc}
    \hline
    \hline
       & $E$ & $\bar{E}$ & $8C_3$ & $8\bar{C}_3$ & $3C_2$,
       $3\bar{C}_2$ & $6S_4$ & $6\bar{S}_4$ & $6\sigma_d$,
       $6\bar{\sigma}_d$ \\
    \hline
    $\Gamma_1$ & 1 & 1 & 1 & 1 & 1 & 1 & 1 & 1 \\
    $\Gamma_2$ & 1 & 1 & 1 & 1 & 1 & $-1$ & $-1$ & $-1$ \\
    $\Gamma_3$ & 2 & 2 & $-1$ & $-1$ & 2 & 0 & 0 & 0 \\
    $\Gamma_4$ & 3 & 3 & 0 & 0 & $-1$ & 1 & 1 & $-1$ \\
    $\Gamma_5$ & 3 & 3 & 0 & 0 & $-1$ & $-1$ & $-1$ & 1 \\
    \hline
    $\Gamma_6$ & 2 & $-2$ & 1 & $-1$ & 0 & $-\sqrt{2}$ & $\sqrt{2}$ & 0 \\
    $\Gamma_7$ & 2 & $-2$ & 1 & $-1$ & 0 & $\sqrt{2}$ & $-\sqrt{2}$ & 0 \\
    $\Gamma_8$ & 4 & $-4$ & $-1$ & 1 & 0 & 0 & 0 & 0 \\
    \hline
    \hline
  \end{tabular}
\end{table}

\begin{table}
  \caption{Character tables of the single valued irreducible representations
  of the point groups $C_{2v}$, $C_{3v}$, and $C_s$.}
  \label{tab:charC2v}
  \centering
  \begin{tabular}[t]{l|cccc}
    \hline
    \hline
    $C_{2v}$ & $E$ & $C_2$ & $\sigma_v$ & $\sigma'_v$ \\
    \hline
    $\Gamma_1$ & 1 & 1 & 1 & 1 \\
    $\Gamma_2$ & 1 & $-1$ & 1 & $-1$ \\
    $\Gamma_3$ & 1 & 1 & $-1$ & $-1$ \\
    $\Gamma_4$ & 1 & $-1$ & $-1$ & 1 \\
    \hline
    \hline
  \end{tabular}
  \hfill
  \begin{tabular}[t]{l|ccc}
    \hline
    \hline
    $C_{3v}$ & $E$ & $2C_3$ & $3\sigma_v$ \\
    \hline
    $\Gamma_1$ & 1 & 1 & 1 \\
    $\Gamma_2$ & 1 & 1 & $-1$ \\
    $\Gamma_3$ & 2 & $-1$ & 0  \\
    \hline
    \hline
  \end{tabular}
  \hfill
  \begin{tabular}[t]{l|cc}
    \hline
    \hline
    $C_s$ & $E$ & $\sigma$ \\
    \hline
    $\Gamma_1$ & 1 & 1 \\
    $\Gamma_2$ & 1 & $-1$ \\
    \hline
    \hline
  \end{tabular}

\end{table}

\begin{table}
  \caption{Character table of the groups $D_{\infty h}$ and $C_{\infty v}$
  (upper left part). These continuous
  groups have four (two for $C_{\infty v}$) one-dimensional representation
  and an infinite number of two-dimensional representations.
  Here $m$ is a positive integer.}
  \label{tab:charDih}
  \centering
  \begin{tabular}{l|ccc|ccc}
    \hline
    \hline
    & $E$ & $2C_\infty^\varphi$ & $\infty\sigma_v$ & $i$ & $2S_\infty^\varphi$ & $\infty C_2$\\
    \hline
    $\Gamma_g^+$ & 1 & 1 & 1 & 1 & 1 & 1\\
    $\Gamma_g^-$ & 1 & 1 & $-1$ & 1 & 1 & $-1$\\
    $\Gamma_g^m$ & 2 & $2\cos m\varphi$ & 0 & 2 & $2\cos m\varphi$ & 0\\
    \cline{1-4}
    $\Gamma_u^+$ & 1 & 1 & \multicolumn{1}{c}{1} & $-1$ & $-1$ & $-1$\\
    $\Gamma_u^-$ & 1 & 1 & \multicolumn{1}{c}{$-1$} & $-1$ & $-1$ & $1$\\
    $\Gamma_u^m$ & 2 & $2\cos m\varphi$ & \multicolumn{1}{c}{0} & -2 & $-2\cos m\varphi$ & 0\\
    \hline
  \end{tabular}
\end{table}

\end{document}